\documentclass[sigconf]{acmart}
\usepackage[normalem]{ulem}
\usepackage{soul}

\usepackage[nolist,smaller]{acronym}

\usepackage{svg}
\usepackage{multirow}
\usepackage{array}
\usepackage{longtable}
\usepackage{supertabular,booktabs}

\DeclareMathOperator*{\codenorm}{norm}
\DeclareMathOperator*{\codevector}{vec}
\newcommand{\Tau}{\mathrm{T}}

\begin{acronym}
    \acro{CUI}{Conversational User Interface}
    \acro{IPA}{Intelligent Personal Assistant}
    \acro{IDE}{Integrated Development Environment}
    \acro{CQA}{Community Question-Answering}
    \acro{API}{Application Programming Interface}
    \acro{RSSE}{Recommendation System in Software Engineering}
    \acro{AJAX}{Asynchronous JavaScript and XML}
    \acro{JSON}{JavaScript Object Notation}
    \acro{HTML}{HyperText Markup Language}
    \acro{URL}{Uniform Resource Locator}
    \acro{VCS}{Version Control System}
    \acro{AI}{Artificial Intelligence}
    \acro{SHOCK}{SHOCK}
    \acro{ROC}{ROC}
    \acro{HCI}{Human-Computer Interaction}
    \acro{SE}{Software Engineering}
    \acro{VR}{Virtual Reality}
    \acro{AR}{Augmented Reality}
    \acro{MR}{Mixed Reality}
    \acro{HMD}{Head-Mounted Display}
    \acro{IE}{Internet Explorer}
    \acro{IEEE}{Institute of Electrical and Electronics}
    \acro{CHI}{the \acs{ACM} Conference on Human Factors in Computing Systems}
    \acro{UIST}{the \acs{ACM} Symposium on User Interface Software and Technology}
    \acro{CSCW}{the \acs{ACM} Conference on Computer-Supported Cooperative Work and Social Computing}
    \acro{VL/HCC}{the \acs{IEEE} Symposium on Visual Languages and Human-Centric Computing}
    \acro{NSF}{National Science Foundation}
    \acro{PI}{PI}
    \acro{co-PI}{co-PI}
    \acro{MOOC}{Massive Open Online Course}
    \acro{IVA}{Intelligent Virtual Assistant}
    \acro{OT}{Operational Transformation}
    \acro{L@S}{The ACM Conference on Learning at Scale}
    \acro{WYSIWIS}{What You See Is What I See}
    \acro{IIS}{Information and Intelligent Systems}
    \acro{VPL}{Visual Programming Language}
    \acro{2D}{two-dimensional}
    \acro{3D}{three-dimensional}
    \acro{CAD}{Computer-Aided Design}
    \acro{FRP}{Functional Reactive Programming}
    \acro{FSM}{Finite State Machine}
    \acro{VRML}{Virtual Reality Modeling Language}
    \acro{UI}{User Interface}
    \acro{GUI}{Graphical User Interface}
    \acro{WIMP}{Window Icon Mouse Pointer}
    \acro{EUP}{End User Programming}
    \acro{CAREER}{CAREER}
    \acro{HTC}{High Tech Computer Corporation}
    \acro{IoT}{Internet of Things}
    \acro{CS}{Computer Science}
    \acro{ACM}{the Association for Computing Machinery}
    \acro{SIGCSE}{the \acs{ACM} Special Interest Group on Computer Science Education}
    \acro{AIED}{Artificial Intelligence in Education}
    \acro{LAK}{Learning Analytics and Knowledge}
    \acro{SI}{School of Information}
    \acro{UMSI}{the University of Michigan School of Information}
    \acro{MOOC}{Massive Open Online Course}
    \acro{AST}{Abstract Syntax Tree}
    \acro{LLM}{Large Language Model}
    \acro{tf-idf}{term frequency–inverse document frequency}
    \acro{T-SNE}{T-Distributed Stochastic Neighbor Embedding}
    \acro{PCA}{Principal Component Analysis}
    \acro{CFG}{Control Flow Graph}
    \acro{ACFG}{Aggregated Congrol Flow Graph}
    \acro{CRDT}{Conflict-free Replicated Data Type}
\end{acronym}
\AtBeginDocument{%
  \providecommand\BibTeX{{%
    \normalfont B\kern-0.5em{\scshape i\kern-0.25em b}\kern-0.8em\TeX}}}

\newcommand{\sys}{{CFlow}{}}

\newcommand{\overcode}{{OverCode}{}}
\newcommand{\runex}{{RunEx}{}}
\newcommand{\vizprog}{{VizProg}{}}
\newcommand{\baseline}{{Baseline}{}}

\newcommand{\add}[1]{#1}
\newcommand{\del}[1]{}

\setcopyright{acmlicensed}
\copyrightyear{2018}
\acmYear{2018}
\acmDOI{XXXXXXX.XXXXXXX}

\acmConference[Conference acronym 'XX]{Make sure to enter the correct
  conference title from your rights confirmation emai}{June 03--05,
  2018}{Woodstock, NY}
\acmISBN{978-1-4503-XXXX-X/18/06}

\begin{document}

\title{\sys{}: Supporting Semantic Flow Analysis of Students' Code in Programming Problems at Scale}


\author{Ashley Ge Zhang}
\orcid{0000-0001-5978-3714}
\affiliation{
    \institution{University of Michigan}
    \city{Ann Arbor}
    \state{Michigan}
    \country{USA}
}
\email{gezh@umich.edu}

\author{Xiaohang Tang}
\orcid{0000-0002-2691-9280}
\affiliation{
    \institution{Virginia Tech}
    \city{Blacksburg}
    \state{Virginia}
    \country{USA}
}
\email{xiaohangtang@vt.edu}

\author{Steve Oney}
\orcid{0000-0002-5823-1499}
\affiliation{
    \institution{University of Michigan}
    \city{Ann Arbor}
    \state{Michigan}
    \country{USA}
}
\email{soney@umich.edu}

\author{Yan Chen}
\orcid{0000-0002-1646-6935}
\affiliation{
    \institution{Virginia Tech}
    \city{Blacksburg}
    \state{Virginia}
    \country{USA}
}
\email{ych@vt.edu}

\renewcommand{\shortauthors}{Zhang, et al.}

\begin{abstract}
% \todo{check word count. up to 350 words. emphasize "at scale" - viewing thousands of solutions simultaneously}
The high demand for computer science education has led to high enrollments, with thousands of students in many introductory courses.
In such large courses, it can be overwhelmingly difficult for instructors to understand class-wide problem-solving patterns or issues, which is crucial for improving instruction and addressing important pedagogical challenges.
In this paper, we propose a technique and system, \emph{\sys{}}, for creating understandable and navigable representations of code at scale.
\sys{} is able to represent thousands of code samples in a visualization that resembles a single code sample.
\sys{} creates scalable code representations by (1) clustering individual statements with similar semantic purposes, (2) presenting clustered statements in a way that maintains semantic relationships between statements, (3) representing the correctness of different variations as a histogram, and (4) allowing users to navigate through solutions interactively using semantic filters.
With a multi-level view design, users can navigate high-level patterns, and low-level implementations.
This is in contrast to prior tools that either limit their focus on isolated statements (and thus discard the surrounding context of those statements) or cluster entire code samples (which can lead to large numbers of clusters---for example, if there are $n$ code features and $m$ implementations of each, there can be $m^n$ clusters).
We evaluated the effectiveness of \sys{} with a comparison study, found participants using \sys{} spent only half the time identifying mistakes and recalled twice as many desired patterns from over 6,000 submissions.

% ORIGINAL:
% In large programming courses, providing personalized feedback to students remains a challenge. With thousands of students in class, it is challenging for instructors to view the submissions simultaneously, understand each student's programming errors, and get insights into their deficiencies and misconceptions. 
% Existing tools that analyze students' errors often limit their focus on outputs or isolated code lines, neglecting the semantic relationships between code statements.
% We introduce \sys{}, a system that enables instructors to identify students' code statement relationship patterns and errors by exploring the semantic flow of thousands of code submissions simultaneously. 
% \sys{} provides multiple viewing levels of the transition from one step to another in thousands of code submissions, helping locate dominant trends in students' solutions. 
% To support pattern discovery, \sys{} semantically aligns and aggregates students' code at a line level, presenting a histogram of similar implementations, while maintaining a formatted code syntax structure for intuitive navigation.
% To facilitate different flow exploration, users can easily switch between different solution approaches using semantic filters.
% With a multi-level view design, users can navigate high-level patterns, and low-level implementations.
% We evaluated the effectiveness of \sys{} with a comparison study, found participants using \sys{} spent only half the time identifying mistakes and recalled more desired patterns from over 6,000 submissions.

\end{abstract}

\begin{teaserfigure}
  \includegraphics[width=\textwidth]{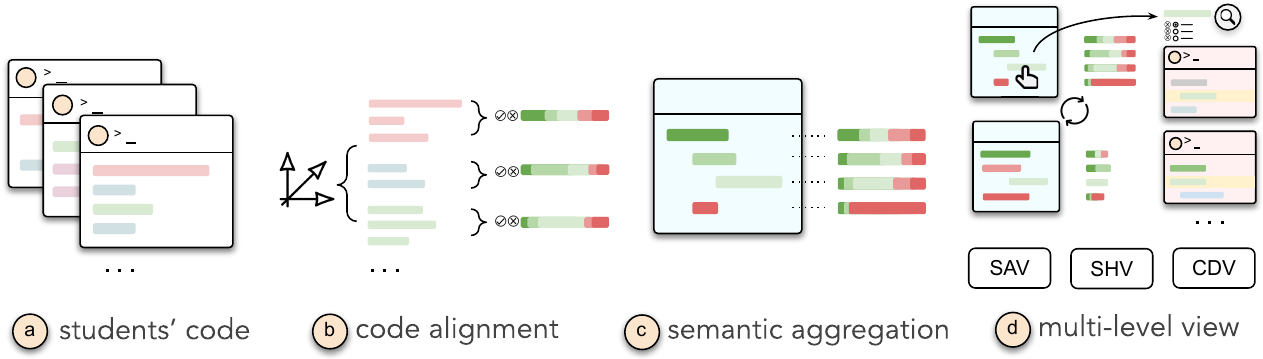}
  \vspace{-0.2in}
  \caption{A system overview of \sys{}. (a) Upon collecting student code submissions, (b) \sys{} extracts and semantically labels the code statements. (c left) These labels are then clustered and represented in a Semantic Aggregation View (SAV). (c right) Statement correctness is color coded and visualized using a stacked bar histogram, forming the Semantic Histogram View (SHV). (d) Users can filter the views by clicking on labels in the SAV or bars in the SHV. The Code Detailed View (CDV) enables for an in-depth inspection of specific code branches.}
  \Description{ .}
  % https://docs.google.com/drawings/d/1T81odpW-9hUiu2SCHsQQq8u0XeN5M0qI7ufIGBZOHUI/edit?usp=sharing
  \label{fig:teaser}
  % \vspace{-0.05in}
\end{teaserfigure}

\begin{CCSXML}
<ccs2012>
 <concept>
  <concept_id>00000000.0000000.0000000</concept_id>
  <concept_desc>Do Not Use This Code, Generate the Correct Terms for Your Paper</concept_desc>
  <concept_significance>500</concept_significance>
 </concept>
 <concept>
  <concept_id>00000000.00000000.00000000</concept_id>
  <concept_desc>Do Not Use This Code, Generate the Correct Terms for Your Paper</concept_desc>
  <concept_significance>300</concept_significance>
 </concept>
 <concept>
  <concept_id>00000000.00000000.00000000</concept_id>
  <concept_desc>Do Not Use This Code, Generate the Correct Terms for Your Paper</concept_desc>
  <concept_significance>100</concept_significance>
 </concept>
 <concept>
  <concept_id>00000000.00000000.00000000</concept_id>
  <concept_desc>Do Not Use This Code, Generate the Correct Terms for Your Paper</concept_desc>
  <concept_significance>100</concept_significance>
 </concept>
</ccs2012>
\end{CCSXML}

\ccsdesc[500]{Do Not Use This Code~Generate the Correct Terms for Your Paper}
\ccsdesc[300]{Do Not Use This Code~Generate the Correct Terms for Your Paper}
\ccsdesc{Do Not Use This Code~Generate the Correct Terms for Your Paper}
\ccsdesc[100]{Do Not Use This Code~Generate the Correct Terms for Your Paper}

\keywords{Do, Not, Us, This, Code, Put, the, Correct, Terms, for,
  Your, Paper}



\settopmatter{printfolios=true}
\maketitle

\section{introduction}
% \todo{introduce "scale" earlier}
Understanding student code is crucial for instructors to provide personalized feedback and enhance student learning outcomes. When performing these tasks, instructors often need to: 1) identify the \textit{issue-relevant code syntax}, and 2) understand the \textit{underlying reasons} for students' struggles. However, with increasing enrollments in CS courses and the diversity of student mental models, these tasks become challenging and tedious for two reasons:
First, a student's code consists of a sequence of code statements (code lines), where each line and the flow from one line to another reflect the student's mental model of the concept. 
This aspect highlights the complexity of understanding code: an issue might be as simple as a one-line error, such as incorrect dictionary usage, or as complex as a mistake spanning multiple lines, involving problems like improper initialization and modification of variable values.
% For instance, an issue might be a simple one-line error, such as incorrect usage of a dictionary, or a more complex mistake spanning multiple lines, involving problems like improper initialization and modification of variable values.
Switching between code syntax and semantics on a large scale is a cognitively demanding task. 
Second, even with identical parts of the syntax, two submissions could yield different outputs, and vice versa. For example, students might arrange their if-else statements differently yet produce the same output. This variation makes it difficult to aggregate a large number of code issues at both the syntax and semantic levels.

Past research has explored this challenge of identifying issues using a variety of methods. 
For example, \overcode{} addressed scalability concerns by clustering and visualizing student code submissions~\cite{glassman2015overcode}. However, they concentrated primarily on specific \del{issues}\add{challenges} such as \del{name errors}\add{different variable names and statement orders,} and focused on the \del{output}\add{computations} of programs, overlooking issues related to \del{thought processes or }high-level misconceptions that might be latent between lines of code. 
To aid in the discovery of diverse high-level patterns, \runex{} enabled instructors to construct runtime and syntax-based search queries with high expressiveness and apply combined filters to code examples~\cite{zhang2023runex}. However, \runex{} depended on instructors to generate these \add{search} queries, which demanded prior knowledge. Additionally, it did not provide guidance on pattern discoverability, leaving users to identify differences between patterns. 
Other auto-feedback generation approaches have been shown to be promising~\cite{wu2021prototransformer}. However, by excluding instructors from the feedback loop, instructors were uninformed about student challenges.
Such detachment could result in instructors not adapting their teaching strategies based on prior student performance. Additionally, there has been evidence suggesting that AI tools can sometimes produce inaccurate results, making one's sole reliance on them risky~\cite{mollick2023using}.

\del{While analyzing code submissions at a large scale, a frequently underestimated, yet crucial, factor is the underlying structure of the code, particularly the interconnections between code statements.}
\add{In this paper, we introduce a novel method that visualizes the flow of code statements to facilitate analyzing students' submissions. Our intuition is that,} similar to the narrative flow in an essay, the sequencing of code statements dictates not only the execution order within a program but also its runtime complexity~\cite{allen1970control}.
Within the context of student submissions, this flow sheds light on a student's grasp of the problem, their approach to problem-solving, thought processes, and potential areas of misunderstanding. 
\del{Concepts such as control flow graphs, data flow graphs, and abstract syntax trees, have been shown to facilitate code analysis. However, aggregating the flow information from thousands of submissions to identify patterns and relationships remains a challenge.}\add{Thus, effectively visualizing this code flow can uncover dimensions of students' understandings that previous research has not addressed.}
% \add{Given that the flow can be broken down into different parts, students can have different implementations and various issues at different parts of the flow. They might share similar misconceptions but have the submissions appear very different or have similar submissions that have distinct misconceptions. This adds to the complexity of understanding the issues within the flow when at scale.}
% Concepts such as control flow graphs, data flow graphs, and abstract syntax trees, have been shown to facilitate code analysis~\cite{allen1970control, kavi1986formal,neamtiu2005understanding}. However, aggregating the flow information from thousands of submissions to identify patterns and relationships remains a challenge.

\add{However, visualizing code flow presents challenges for two reasons: First, student code submissions often vary significantly, not just in their solutions but also in their structure and semantic meaning. The difficulty lies in aggregating and aligning these diverse submissions. 
Second, designing a visualization that preserves the code's inherent structure while also enabling analysis of its flow is complex, as different code implementations may correspond to the same flow. 
Moreover, aggregating code structures on a large scale could make it more cognitively demanding for instructors to interpret the information. The visualization needs to present flow information in a manner that aligns with how people typically read and comprehend code.}

\del{Specifically, these challenges include: First, with the \textbf{Code Alignment and Semantic Variance}. Students often employ varied problem-solving methods, resulting in diverse implementations. The challenge is not only in aggregating these varied solutions, but also in aligning code that differs in structure and semantic meaning. Second, with the \textbf{Visualization of Diverse Flow Patterns}. Tools like \overcode{} and \vizprog{}  cluster code based on similarities , however they struggle with the diverse flows that reflect students' mental models. Thus, designing a visualization that maintains the code's inherent structure while facilitating flow analysis is complicated. }

\add{To tackle these challenges, we designed \sys{}, a system comprising three distinct views (Figure~\ref{fig:teaser}d): the \textit{Semantic Aggregation View} (SAV), the \textit{Semantic Histogram View} (SHV), and the \textit{Code Detailed View} (CDV). 
In essence, \sys{} aims to represent semantic flows between distinct value sets and showcases categorical variances. 
Specifically, \sys{} employs a multi-step algorithm where each code line is subjected to a detailed semantic analysis and error check using \ac{LLM}. These lines are then vectorized using CodeBert to group them based on similarity. A “common progression” of steps from correct solutions provides a reference structure for mapping all solutions. The resultant structured data is then visualized in a color-coded format, enabling educators to quickly pinpoint student challenges and misconceptions. This method 1) highlights semantic patterns by frequency and accuracy, and 2) simplifies the navigation and comparison of code flows. The core insight of \sys{} is to align instructors' analysis of code submissions with the intrinsic characteristics of student code.}

To assess the efficacy of \sys{} in aiding instructors in identifying issues with student code, we conducted a within-subject experiment involving 16 participants and over 6,000 student code submissions for two programming exercises. 
\add{To ensure a fair comparison, we designed the baseline system to be the combination of two state-of-the-art systems, \overcode{} and \runex{}~\cite{glassman2015overcode, zhang2023runex}.}
Our findings indicated that, in comparison to \del{a}\add{the} baseline system, \sys{} enabled participants to 1) identify targeted misconceptions in half the time used for the baseline, and 2) achieve greater accuracy in their results. 
\add{\sys{} is the first system that bridges the high-level flow among thousands of submissions with specific student errors. By overlaying aggregated semantic data on submissions while retaining their inherent structure, \sys{} enhances exploration and navigation capabilities.}

% Moreover, participants noted that \sys{} facilitated an easier exploration of patterns in students' code and enhanced the comprehension of code structure information. 
Our research underscores the continued need to assist instructors in analyzing large-scale, structurally intricate student data.
This research thus contributes:

\begin{itemize}
    \item \add{A novel visualization approach, and implementation of the approach, that aggregate semantic patterns and code structures, to unearth complex in large-scale, multifaceted code submissions.}
    \item Evidence from an evaluation of \sys{} that suggests that \sys{} can assist users in identifying a variety of patterns and misconceptions in students' code.
\end{itemize}

\section{related work}
Our work builds on prior work on programming education at scale, code flow analysis, and LLM for programming education.

\subsection{Mistakes in Introductory Programming}

There are primarily three categories of programming knowledge in introductory courses: 1) syntactic knowledge, such as language features and rules, 2) conceptual knowledge, such how programming constructs and concepts work 3) strategic knowledge, which refers to how to apply prior knowledge to solve programming problems~\cite{bayman1988using, mcgill1997conceptual, qian2017students}.
Syntactic mistakes are frequent, but are usually superficial and easy to fix~\cite{altadmri201537, jackson2005identifying}. Conceptual mistakes could lead to significant misconceptions and are relevant to students' thought process~\cite{qian2017students, bayman1983diagnosis, canas1994mental, ma2007investigating, sorva2012visual}. For instance, errors on variable initialization and modification relate to various misconceptions, such as variable scopes~\cite{fleury1991parameter} and where they are stored~\cite{bayman1983diagnosis}. Furthermore, errors on concepts like conditionals and looping constructs can lead to misconceptions on program execution~\cite{green1977conditional, sirkia2012recognizing, sleeman1986pascal}. Students that lack syntactic and conceptual knowledge could make more strategic mistakes~\cite{ebrahimi1994novice, de2008teaching}, reflecting difficulties in decomposing the programming problem~\cite{muller2005pattern, robins2006problem}. To identify different types of misconceptions, tools should reveal the various dimensions of students' code. \textbf{Therefore, our first design goal (DG1) is to easily understand the semantic flow within students' code.}

\subsection{Programming Education at Scale}
% \todo{introduce concepts from learning at scale }
Large courses, such as Massive Open Online Courses (MOOCs), face challenges of maintaining quality and offering personal attention to learners~\cite{hew2014students}. 
In the context of programming education, the challenge lies in analyzing students' coding submissions to understand their learning needs and provide tailored feedback. However, the expansive course size and the wide variation among students' coding solutions make it submissions time-consuming and laborious~\cite{glassman2015overcode, guo2015codeopticon, wang2021puzzleme, chen2020edcode, chen2021towards, medeiros2018systematic, head2017writing}.

To address these challenges, various tools have been crafted to provide instructors with an overview of students' code~\cite{glassman2015overcode, guo2015codeopticon, zhang2023vizprog, gaudencio2014can, huang2013syntactic}. 
However, these tools take code submission as single piece and ignore the multifaceted aspect of the program flow, hindering instructors' ability to analyze the flow within thousands of code samples simultaneously.
Code search tools could help instructors identify specific coding patterns~\cite{nguyen2014codewebs, zhang2023runex, podgurski1993retrieving, mathew2020slacc, mathew2021cross}. Codewebs was a code search engine in educational settings that employed Abstract Syntax Trees (ASTs) and unit test outcomes to match code samples, enabling instructors to index a million code submissions~\cite{nguyen2014codewebs}. Codewebs' was limited to filtering by post-execution runtime values. 
\runex{} allowed instructors to construct queries based on runtime and syntax with high expressiveness and to search by combined filters~\cite{zhang2023runex}. 
The downside of code search tool is that instructors need to create queries manually, requiring knowledge of student approaches and challenges. In contrast, \sys{} reduces the effort needed for query formulation by analyzing and summarizing code flows across submissions and offering an ``available'' vocabulary.

\subsection{Code Flow Analysis in CS Education}

Control flow, or the interrelation of statements in a program, is crucial to programming comprehension ~\cite{allen1970control}, providing a lens into the programmer's logic, strategy, and misconceptions.
Understanding code flow becomes paramount at scale due to the variety of coding solutions from students with diverse backgrounds and thought processes, posing challenges in assessing the correctness of a solution, discerning approaches, identifying misconceptions, and offering tailored feedback~\cite{keuning2019teachers, singh2013automated, kaczmarczyk2010identifying}.

Current educational methods emphasize the process of coding over product, focusing on student crafting their solution as seen in their code flow~\cite{taniguchi2022visualizing}.
Tools like Theseus~\cite{lieber2014addressing} visualize code segment interactions to spot logical errors. However, their designs struggle with numerous varied solutions in large CS courses.

\sys{} overcame this challenge by offering multi-level views of students' solutions with semantic labeling and interactive filtering. 
\sys{} combined elements from Sankey diagrams, adapted for mapping code structures and logic pathways, and histograms for highlighting data distribution and identifying patterns in student code~\cite{riehmann2005interactive,scott1979optimal}.
The combined utility of these visualization techniques in a cohesive system remains largely unexplored in programming education. \sys{}'s multi-faceted visualization bridges this gap by integrating the flow-centric insights of Sankey diagrams with the succinct data representation of histograms. \sys{} simplifies Sankey diagrams to align with how people comprehend code, while preserving interactive flow relationships.
\sys{} offers a scalable and intuitive method for instructors to navigate numerous student submissions and discern patterns in large CS courses.

\subsection{LLMs for Programming Education}
Large language models (LLMs)~\cite{metz2021ai} have been widely used in programming education for code generation~\cite{kiesler2023large, eckerdal2005novice}, program comprehension~\cite{pankiewicz2023large}, language learning~\cite{becker2023programming}, and teaching material design~\cite{lu2023readingquizmaker}.
Prior work has designed systems using LLMs to help instructors comprehend students' code more effectively than student-generated explanations~\cite{zhang2023vizprog, leinonen2023comparing, hellas2023exploring, leinonen2023comparing}. 
% Compared to code explanation generated by students, LLM-created explanations are easier to understand and more accurate summaries of code~\cite{}.
However, LLMs face challenges in conveying complex relationships and structures in code at scale due to their text-based nature.~\cite{chang2002effect}. 
To tackle the issue, \vizprog{} visualizes students' coding progress as dynamic dots on a 2D map in real time, through clustering vector embeddings of students' solutions using pre-trained language model CodeBERT~\cite{zhang2023vizprog}.  
\vizprog{}'s downside is that it only one level of abstraction, losing structural cues in syntax and semantic meanings. 
To enhance understanding, a progressive disclosure approach is suggested to gradually provide pieces of information about code that contribute to the overall understanding~\cite{pea1987user}. 
\textbf{Therefore, our second design goal (DG2) is to seamlessly navigate students' code between high level abstraction and detailed information.}
\sys{} builds on this by using LLM-created content to aid instructors in processing students' code at scale, employing a hierarchical and ``focus + context'' design to visualize numerous text information.   

\section{\sys{}}
\subsection{System Design Goals}
Derived from the prior literature, we developed three design goals (DG1--DG2) to guide the development of \sys{} to help instructors understand and explore student solutions to a programming exercise.

\begin{itemize}

    \item \textbf{DG1: Easily understand the semantic flow within students' code.} Comprehending the semantic flow is essential for grasping code's structure and design. The system should provide an intuitive, concise view that simplifies the semantic flow analysis of students' solutions, thus enhancing overall code comprehension. Visualization of code flow at large scale could be difficult to digest, thus the visualization should align with how human read and comprehend code.
    
    \item \textbf{DG2: Seamlessly navigate students' code between high level abstraction and detailed information.} When reviewing students' code on a large scale, it's crucial for instructors to constantly dive into specific submissions to ground their understanding. This allows instructors to identify concrete examples and form contextualized feedback. To support this behavior, the system should offer easy navigation on students' code, thus enabling instructors to switch between ``diving in'' and ``floating up'' different samples.

    % \item \add{\textbf{DG1: Easily view the multifaceted program flow.}}
    % \item \add{\textbf{DG2: Aggregate large-scale code flow in a concise view.}}
    % \item \add{\textbf{DG3: Visualize in line with how human read and process code.}}
\end{itemize}

\add{With these design goals in mind, \sys{}'s visualization integrates the flow-centric insights of Sankey diagrams with the data distribution conciseness of histograms to present students' code flow. This design can effectively illustrate how data or control moves through different parts of the program. Combined with histograms, the visualization could provide instructors with an overview and reveal bottlenecks in the flow. Furthermore, \sys{}}
\del{With these design goals in mind, \sys{}'s visualization} provides instructors with a multi-level view of their students' code, thus enhancing their ability to navigate and explore code at scale. It empowers them to discern patterns and identify mistakes, all within the contextual framework of semantic flow analysis. \add{We implemented \sys{} as a prototype.}
In the following sections, we describe \sys{}'s user interface and the algorithms that were developed to support its features. 

\subsection{\sys{}'s User Interface}

\sys{}'s user interface is organized into three primary panels: a  Semantic Aggregation View (SAV; Figure ~\ref{fig:system}a), a Semantic Histogram View (SHV; Figure ~\ref{fig:system}b), and a Code Detailed View (CDV; Figure ~\ref{fig:system}e).
Upon loading student code submissions, \sys{}'s algorithm (described in detail in Section~\ref{sssec:algorithm}) determines the semantic meaning and error information of each code line across all submissions, cluster code lines by their vector similarity, and determined the correctness of the implemented solutions. 
% The SAV displayed code syntax that was aligned based on the semantic similarity between individual lines across all submissions, with color-coding to represent the correctness of each line. 
The SAV displayed code syntax that align with a common semantic flow among all submissions, with color-coding to represent the correctness of each step in the semantic flow.
% In contrast, the SHV offered a historical view of all code that had semantic similarities to a reference code line, again using color to denote correctness.
In contrast, the SHV offered a historical view of all code implementations that were clustered as the same step in the SAV, again using color to denote correctness.
% When a user clicked on any of these lines, the CDV updated its view, showcasing a curated list of code submissions that contained the selected line. Concurrently, both the SAV and SHV updated to reflect this user interaction.
When a user clicks any of these code lines, the CDV updated its view, showcasing a curated list of code submissions that contained the selected line. Concurrently, both the SAV and SHV updated to reflect this user interaction.
In the following sections, we illustrate the details of the system's design.

\subsubsection{The SAV: Semantic Reference Code}
% To ease pattern identification (DG1), \sys{} presented users with a familiar code structure (Figure ~\ref{fig:system}a). Contrary to initial impressions, the code displayed here was not sourced from a particular submission. Instead, each line was a sample that denoted a label or a category, which might originate from different student submissions. 

\begin{figure*}
    \centering
    \includegraphics[width=\textwidth]{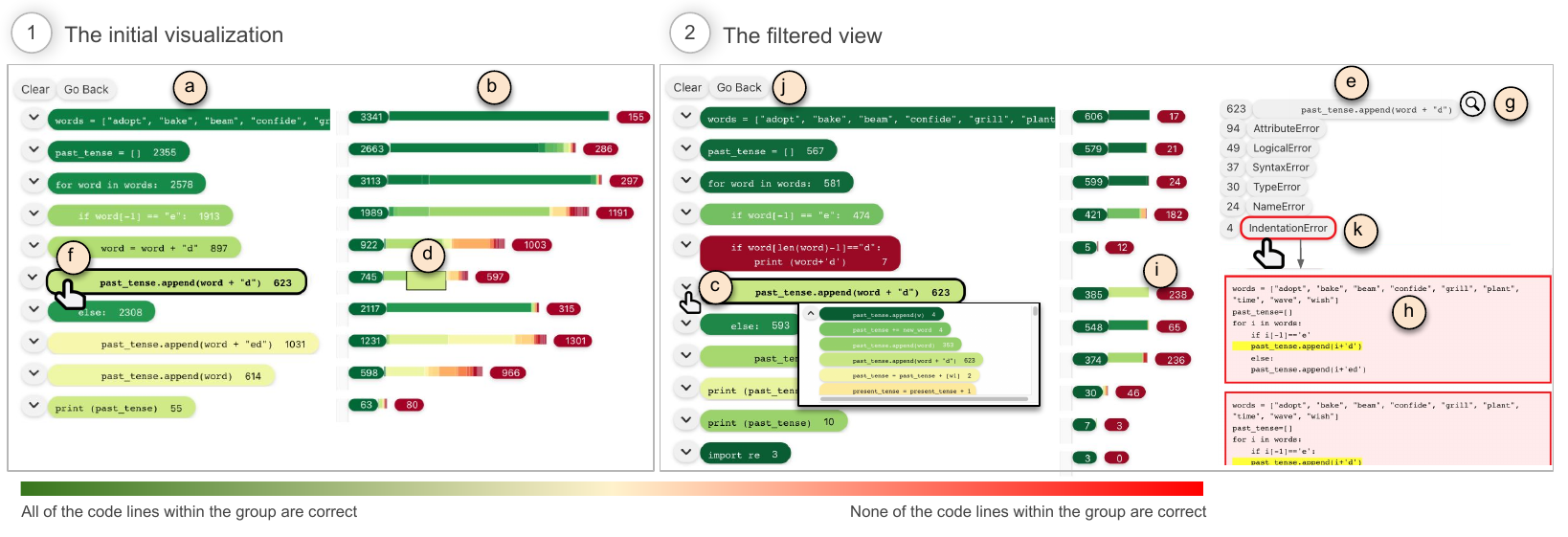}
    \vspace{-0.2in}
    \caption{\sys{} allows users to explore the semantic flows of student code submissions at a high level through semantic aggregation. First, users are presented with an overview of the entire set of solutions (1), including the SAV (a) and the SHV (b). They can then click on individual code lines (f) to progressively explore the details of specific implementations. The visualization will update to focus on a smaller subset of solutions, displaying the aggregated flow and distributions of the selected set only (2). Users can inspect details of the flow at individual code level in the CDV (e). \sys{} offers a breakdown of types of errors within that group (k), and detailed solutions with context-aware highlighting (h).}
    \label{fig:system}
    \vspace{-0.2in}
    % https://docs.google.com/drawings/d/1DuMKlGfxI7Gu-Gg9v05M2EVuNV6gW44y2nDXDTcLgI4/edit?usp=sharing
\end{figure*}

To ease pattern identification (DG1), \sys{} presents code in a visualization that is resembles a single code sample (Figure ~\ref{fig:system}a). Nevertheless, this visualization did not originate from a single submission; instead, each line symbolizes a group of similar code lines, like a label or category, drawn from different student submissions. These lines are arranged in the identical sequence as they were found in the students' work. For instance, there was a coherent progression from initializing variables, iterating over variables, to employing conditional statements, mirroring the typical structural approach employed by most students in their code.

Different from conventional views where correctness is about the whole solution, here we display correctness at code line level. Each line's color-coded representation denotes the correctness of the associated group of code lines. The colors were mapped on a scale from [0, 1] to [Red, Green]. In this scale, Red signified that none of the code lines within the group were correct, whereas Green indicated that all the code lines within the group were correct.

% \begin{figure*}
%     \centering
%     \includegraphics[width=\textwidth]{images/cflow_system_2.pdf}
%     \caption{Users can inspect the details of each flow at the individual code level (e) in the filtered view (a). For each specific group, \sys{} offers a breakdown of types of errors found in the code lines within that group (f). By clicking on a particular error type, users can access detailed solutions that feature context-aware highlighting (h). Additionally, by clicking the expand button (b), users can reveal a dropdown list displaying all the code lines associated with a specific group (c).}
%     \label{fig:system}
%     % https://docs.google.com/drawings/d/1DuMKlGfxI7Gu-Gg9v05M2EVuNV6gW44y2nDXDTcLgI4/edit?usp=sharing
% \end{figure*}

To facilitate user exploration of the semantic flow (DG2), \sys{} implements clickable elements for all components within the SAV. Users have the option to click on the expand button located to the left of each code line (Figure~\ref{fig:system}c), revealing similar labels (Figure~\ref{fig:system}c) that correspond to the same position in the reference code framework. Alternatively, users can directly click on the code line itself \add{(Figure~\ref{fig:system}f)}. This action simultaneously updates all system views \add{from the initial visualization of the entire solution set (Figure~\ref{fig:system}.1) to the filtered view of the subset selected (Figure~\ref{fig:system}.2)}. To enhance navigation, users have the option to click the `Go Back' button to return to a broader view (Figure~\ref{fig:system}j). The `Clear' button, on the other hand, allows users to instantly revert to the most expansive view level.

% To make it easy for users to explore the flow (DG2), \sys{} made all components in SAG clickable. Users can click on the expand button left to each code line (Figure ~\ref{fig:system}b), thus revealing similar labels (Figure ~\ref{fig:system}c) that maintained the same position in the reference code. Users can also click on the code line itself. This action concurrently updated all system views, filtering the displayed content to submissions that included the selected line (Figure ~\ref{fig:system}h, i, f). Additionally, a comprehensive list of related code submissions was displayed on the CDV which we will introduce next (Figure ~\ref{fig:system}.f), granting users access to detailed, low-level information. An automatically generated list of error types regarding the selected code line further aided users, acting as filters to streamline the search process (Figure ~\ref{fig:system}j).
% To enhance navigation, users could click the `Go Back' button to revert to a broader view (Figure ~\ref{fig:system}g). The `Clear' button instantly returned users to the most expansive view level.

\subsubsection{The SHV: Semantic Code Histogram \& Line Correctness}
% new version
To assist users in identifying multiple patterns within the data (DG1), \sys{} plots a histogram based on the semantic labels detailed in the previous section. The color-coding within this view served as an indicator of a collection of code line's correctness. Each stacked bar view is flanked by two numbers representing the count of code lines that are correct and incorrect, respectively.
We incorporated the correctness metrics to cater to the needs of users who might be  interested in submissions that had an error due to a particular line or chunk of code. An  animation (Figure ~\ref{fig:system}d) emphasized the proportion of the current highlighted labels within the overall dataset, thus enhancing navigational ease.
Interacting with the SHV was designed to be consistent with the SAV, so clicking on any bar within the SHV triggered a system response just like selecting a label did in the SAV (Figure ~\ref{fig:system}f). Consequently, the SHV (Figure ~\ref{fig:system}i) retained only the clicked bar, while the rest of the histograms adjusted based on their correctness metrics.
This interaction paradigm facilitated a nested querying approach. By using a label as a filter, users can delve deeper into specific data subsets, mirroring the nested query functionality in database searches. Here, sequential queries systematically refined the search path, guiding users toward their desired data branch.

% original version
% To assist users in comparing multiple patterns within the data (DG1), \sys{} plotted a histogram based on the semantic labels detailed in the previous section. The color-coding within this view served as an indicator of a line's correctness. Each stacked bar view was flanked by two numbers representing the count of correct and incorrect submissions, respectively.
% We incorporated these dual correctness metrics to cater to the needs of users who might be  interested in submissions that had an error due to a particular line or chunk of code. An  animation (Figure ~\ref{fig:system}d) emphasized the proportion of the current highlighted labels within the overall dataset, thus enhancing navigational ease.
% Interacting with the SHV was designed to be consistent with the SAV, so clicking on any bar within the SHV triggered a system response just like selecting a label did in the SAV (Figure ~\ref{fig:system}b). Consequently, the SHV (Figure ~\ref{fig:system}i) retained only the clicked bar, while the rest of the bars adjusted based on their correctness metrics.
% This interaction paradigm facilitated a nested querying approach. By using a label as a filter, users could delve deeper into specific data subsets, mirroring the nested query functionality in database searches. Here, sequential queries systematically refined the search path, guiding users toward their desired data branch.

\subsubsection{The CDV: Code Search with Error Filters \& Semantic Labels}
% new version
To assist users in quickly navigating and filtering through students' code by different categories such as error types or specific semantics (DG2), \sys{} contains a code search engine equipped with features such as filters (Figure~\ref{fig:system}g) and context-aware code syntax highlighting (Figure~\ref{fig:system}h). What sets this design apart from other code search utilities, such as \runex{}, is the intricate data-binding that existed across the three primary views of \sys{}.
As previously discussed, users can initiate a search by clicking on a label or a histogram bar, with the clicked label serving as the search query. This interaction enabled users to continually refine their search with additional click-based queries and layer multiple filters, thus progressively narrowing the list of relevant code submissions.

After uploading code submissions, error type filters were auto-generated using a \ac{LLM} (as section \ref{sssec:algorithm} explains in further detail). The accompanying numbers (Figure~\ref{fig:system}k) designates the count of submissions that exhibited the corresponding error type at the selected code line. For example, in the scenario depicted in Figure~\ref{fig:system}e, out of the 623 submissions containing the \texttt{past\_tense.append(word + ``d'')} line, over 200 submissions were identified as having an error at this specific line by the \ac{LLM}, while the remainder were deemed correct at that line. Users can seamlessly navigate this list to inspect each relevant submission in detail.

\subsection{\sys{}'s Algorithm}\label{sssec:algorithm}

\begin{figure*}
    \centering
    \includegraphics[width=\textwidth]{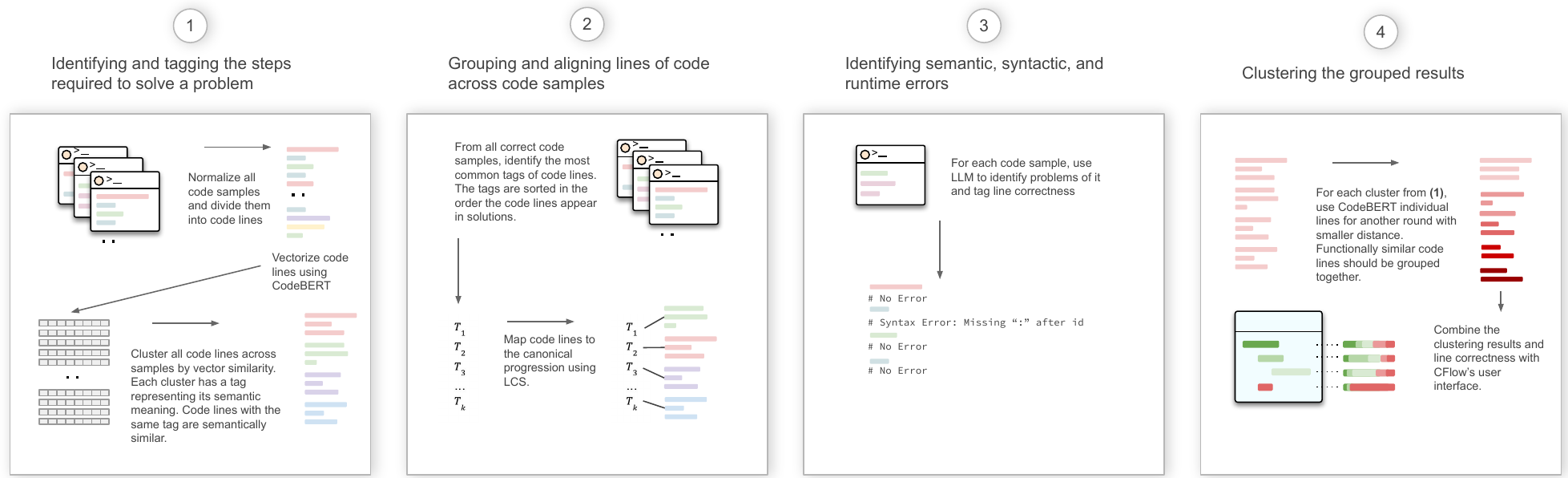}
    \vspace{-0.2in}
    \caption{\add{\sys{}'s algorithm. To generate the results required in \sys{}'s user interface, \sys{}'s algorithm include four primary stages: (1) identifying and tagging the steps required to solve a problem, (2) grouping and aligning lines of code across code samples, (3) identifying semantic, syntactic, and runtime errors, and (4) clustering the grouped results. }}
    \label{fig:algorithm}
    \vspace{-0.2in}
    % https://docs.google.com/drawings/d/10B93zNMDXfN_uCJI8Ex8aof2Up7tGtXJRGXhbx7l_qE/edit?usp=sharing
\end{figure*}

\add{To generate the aforementioned visualization, \sys{} requires detailed information about correctness of students' code at the line level. To effectively visualize the information, \sys{} segments the code into various components based on the semantic flow and then group these segments according to similarity. Specifically, }
\sys{} produces its visualization using four primary stages \add{(Figure \ref{fig:algorithm})}:

\begin{itemize}

    \item \textbf{Stage 1:} Identifying and tagging the steps required to solve a problem
    \item \textbf{Stage 2:} Grouping and aligning lines of code across code samples
    \item \textbf{Stage 3:} Identifying semantic, syntactic, and runtime errors
    \item \textbf{Stage 4:} Clustering the grouped results
\end{itemize}

We will describe each stage in the sub-subsections below.

\subsubsection{Stage 1: Identifying and tagging the steps required to solve a problem}
First, \sys{} identifies the common steps across the code samples (Figure \ref{fig:algorithm}.1).
Doing this requires understanding the \emph{semantic meaning} of these code samples.
For example, ``\texttt{if x != 5:}'' is functionally identical to ``\texttt{if not (2+3 == i):}'' if \texttt{x} and \texttt{i} serve the same purpose in code.
Further, many code samples contain minor syntax errors (like omitting a parentheses in ``\texttt{if not 2+3 == i):}'') that should be classified as semantically similar.

Thus, for each code sample (which we will denote as $c_i$ to represent code sample $i$) \sys{} first `normalizes' the code using text-based normalization techniques.
This includes removing extra whitespace, stripping code comments, identifying variable names and remapping them to be consistent across code samples, and removing \texttt{print()} function calls.
We will refer to the `normalized' version of code sample $c_i$ as ``$\codenorm{}(c_i)$''.
\sys{} then divides the normalized code sample into lines; we will denote the first line of $c_i$ as $c_i^1$, the second line as $c_i^2$, etc.
\sys{} then uses CodeBERT~\cite{feng2020codebert} to vectorize  of $\codenorm{}(c_i)$, which we will denote as $\codevector(\codenorm(c_i)^j) \in{} \mathbb{R}^{768}$ for line $j$ of code sample $i$.
We will use $v_i^j$ as shorthand for $\codevector(\codenorm(c_i)^j)$.
This vectorized representation ($v_i^j$) captures the \emph{semantic} meaning of the code line and is resilient to small variations and errors.
It is also contextualized in the larger code sample to capture what $\codenorm(c_i)^j$ means \emph{in context}.

Next, \sys{} clusters all code lines across samples $\{v_1^1, v_1^2, ..., v_2^1, v_2^2, ...\}$ to group similar lines of code.
Each line of code is placed into one cluster.
We will use $\tau{}_i^j$ to represent the ID of the cluster that line $v_i^j$ is placed in.
The cluster ID $\tau{}_i^j$ is then used as a \emph{tag} for each line.
At the conclusion of stage 1, each line of code in each code sample has a tag ($\tau{}_i^j$) that represents its semantic meaning.
Every line of code with the same tag $\tau{}$ is semantically similar.

\subsubsection{Stage 2: Grouping and aligning lines of code across code samples}
In stage 1, \sys{} created semantically meaningful labels ($\tau{}_i^j$ is the label for line $j$ of code sample $i$).
Although these labels capture the semantic meaning of every given code line, they were created independent of each line's location in the larger code sample, a necessary component for \sys{}'s visualization to represent many code samples coherently.
To do this \sys{} then \emph{aligns} lines of code across samples \add{(Figure \ref{fig:algorithm}.2)}.

As a first step for alignment, \sys{} identifies a ``canonical progression''---a set of steps that represents the ``average'' correct solution.
To identify these steps, \sys{} first identifies which code samples are \emph{correct}, using unit tests.
Across each \emph{correct} solution $\codenorm(c_i)$, \sys{} identifies the most common tag for each individual line across code samples.
We denote this progression as $\Tau{} = (\Tau{}_1, \Tau{}_2, ...)$.
This means, $\Tau{}_1$ is the most common value of the tag for the first line ($\tau{}_i^1$) among correct solutions, $\Tau{}_2$ is the most common value for $\tau{}_i^2$, etc.
The length of the canonical progression $\Tau{}$ depends on a pre-set parameter that determines the minimum agreement across solutions, stopping once it reaches a line number where tags are not sufficiently consistent.
Finally, \sys{} computes the longest common subsequence (LCS) between each code sample's tags $(\tau{}_i^1, \tau{}_i^2, ...)$ and $(\Tau{}_1, \Tau{}_2, ...)$.

One assumption made by this approach is that there is one representative ``average'' progression $\Tau{}$---that most correct solutions use a similar set of steps.
Future iterations of this algorithm could address this limitation by producing a value of $\Tau$ for each approach.

\subsubsection{Stage 3: Identifying semantic, syntactic, and runtime errors}\label{sec:LLM-error}
An important aspect of \sys{}'s visualization is that it helps users understand \emph{common problems} and errors that learners face.
Testing code samples against unit tests can identify which code samples contain errors but does not help identify which specific parts of the code samples are problematic.
Further, even identifying lines that result in runtime or syntax errors might not help identify the true actual source of a given error.
For example, a runtime error might occur on a correct line if it references a variable that was incorrectly set on a line before it.

Thus, to identify lines of code with errors with more precision, \sys{} passes each code sample $c_i$ through an LLM (ChatGPT-3.5) that is prompted to identify problems in the code \add{(Figure \ref{fig:algorithm}.3)}.
For each line, this LLM identifies whether the line is (1) correct or contains (2) a semantic error, (3) a syntax error, or (4) a runtime error.
We denote this information for line $j$ of code sample $i$ as $E_i^j$.
The correctness at the code line level was determined using GPT3.5-turbo. The prompt we used for this determination can be found in appendix (Section \ref{sec:prompt}).

% \soney{move prompt to appendix}

\subsubsection{Stage 4: Clustering the Grouped Results}
After \sys{} identifies a canonical progression ($\Tau{}$) and maps each line across code samples to that progression (stage 2), it then clusters every lines of code within each tag $\tau{}$ (Figure \ref{fig:algorithm}.4).
Recall that the tag for line $j$ of code sample $i$ ($\tau{}_i^j$) was computed by clustering vectorized code lines ($v_i^j$).
As a result, code lines that are functionally identical (such as ``\texttt{if x != 5:}'' and ``\texttt{if not 2+3 == i):}'') should end up with the same tag $\tau{}$.
Finally, for every line, \sys{} then performs another round of clustering of code line vectors $v$ \emph{within} grouped lines with the same tag $\tau{}$, with a smaller maximum distance metric.
These clusters are then used to group solutions within the \sys{} UI and are combined with the error information generated in step 3.

\add{\subsection{Evaluation of \sys{}'s Algorithm}
Using LLM-powered approaches could raise certain concerns, including low transparency, lack of control, and issues of trust~\cite{liao2023ai}. To determine if the performance of the LLM impacts \sys{}'s effectiveness, we conducted a comparative study assessing the accuracy and reliability of our approach, against human judgment. 
Specifically, we analyzed how well the LLM (GPT-3.5-turbo\footnote{\url{https://platform.openai.com/docs/models/gpt-3-5}}) used in \sys{} identifies issues within code samples (details in Section \ref{sec:LLM-error}). We did not evaluate the other LLM-powered component in \sys{}, CodeBERT, as its efficacy has been extensively studied and validated in previous research\cite{feng2020codebert, mashhadi2021applying, zhou2021assessing}.}

\sys{} used LLM to label each line of a code sample as either (1) correct, (2) having semantic error(s), (3) having syntax error(s), or (4) having runtime error(s). 
We evaluated the LLM's performance in identifying the semantic errors, by comparing that of expert human labeler. 
However, expert human labeler could have different opinions on which lines are incorrect in a submission. Some people would only take lines that need edits as incorrect, while other people would take the whole program construct involving the erroneous lines as incorrect. Instead of designing a rubric of a single standard and label ground truth, we recruited two experienced Python programmer to label 50 solutions to 5 programming exercises, selected code lines that are incorrect, and compared the results to LLM's outputs. All the solutions in the dataset had semantic errors.

In the study, the first participant (L1) annotated the dataset by marking the code lines with issues. L1 was then asked to review the LLM's annotations and update their original labels if necessary. The second participant (L2) received both L1's revised annotations and the LLM's annotations, but was not informed about the source of each label set. L2's task involved comparing these two annotation sets and choosing one of four options: (1) version 1 is correct, (2) version 2 is correct, (3) both versions are correct, or (4) both versions are incorrect. Beyond evaluating LLM's performance, this study design accounts for potential biases in human labeling (e.g., level of specificity) and enables us to evaluate the representativeness of the LLM's labels (e.g., whether the incorrect labels are made up).

To evaluate the level of agreement, we considered labels marking specific lines with isseus and labels applied to the entire code block containing that line as equivalent. This approach is justified as human labelers might annotate lines with issues at varying levels of specificity, yet still indicate the same error.
For instance, in a solution with an incorrect combination of condition and content in an if-else statement, labeling just the condition and labeling on the entire if-else statement were considered the same. Our analysis of the labeling results showed that the two participants agreed on 96\% of the lines. L1 incorrectly labeled two solutions, which L2 subsequently corrected. LLM's outputs agreed with L1 and L2 on 80\% and 90\% of the lines, respectively. However, upon closer examination of the specific solutions and error messages generated by the LLM, we discovered that LLM actually achieved 96\% and complete agreement with L1 and L2, respectively. For example, when incorrect solutions missed variables that are required by the exercise, L1 labeled the lines where the variables were initialized, whereas LLM labeled the lines where the variables were used. Therefore, we believe that \sys{}'s algorithm is reliable for educational settings. We will discuss this with more details in Section \ref{sec:llm-discussion}.

% \todo{(1) percentage (2) fundamental agreement on errors (3) errors}
\section{user study}
A within-subject study was conducted to evaluate \sys{}'s efficacy in supporting instructors to understand student code submissions at scale. 
The baseline tool, combined the core functionalities of two state-of-the-art research tools, \runex{} and \overcode{}, that were designed with a similar goal to \sys{}~\cite{zhang2023runex, glassman2015overcode}. 
In the study, participants used either the baseline system or \sys{} to answer quiz questions on students' errors on patterns and logic consistency.
The Institutional Review Board (IRB) approved the study, ensuring adherence to ethical standards and participant safety. 

\subsection{Method}
\subsubsection{Participants}

Because \sys{}'s end users would be programming instructors, we reached out to senior students from \textit{[redacted for anonymity]} who had experience teaching Python programming courses. During a screening session, participants indicated their prior experience teaching and using Python. Teaching assistants and senior students with at least 3 years of experience were invited to participate in the study. \add{Given that participants with teaching experience might anticipate certain student mistakes, we also included non-teaching participants that are senior students and experienced in Python.}
In total 16 participants were recruited (i.e., 4 self-identified as male, 12 as female) and their experience with Python programming ranged from 2 years to 8 years, with 14 participants having previously taught programming courses in Python and the other 2 being senior students that are experienced in Python.

\subsubsection{Study Systems}
% \gez{make this subsection more concise}

% 

% borrow design goals from RunEx and OverCode to justify why we pick these tools. They are designed for ... purposes and evaluated by user studies ...
Since there is no widely used commercial tools that identify students' mistakes and approaches in code to compare with, we designed the baseline system as a combination of \runex{}~\cite{zhang2023runex} and \overcode{}~\cite{glassman2015overcode}. Both \runex{} and \overcode{} are the-state-of-art systems for viewing students solutions. Their user studies showed benefits from various aspects - \runex{} helps users identify specific code patterns with higher accuracy and expressiveness~\cite{zhang2023runex}, while \overcode{} allows teachers to quickly develop a high-level view of students' understanding and misconceptions, and to provide relevant feedback to students~\cite{glassman2015overcode}.
However, \overcode{}'s limitation was in analyzing mistakes within students' code, as it required solutions to be free of syntax errors, and categorized solutions that were syntax-error-free but failed the unit tests in their own distinct cluster. 
To fill this missing part of \overcode{} and ensure a fair comparison, we complemented \overcode{}'s clustering results with \runex{}~\cite{zhang2023runex}, a code search tool designed for programming education. \runex{} enables users to explore class-wide patterns within a large volume of student code by augmenting regular expressions with runtime values for enhanced functionality. During the user study, we used the user interface of \runex{} and display \overcode{}'s clustering results in it. Both \sys{} and the baseline systems were implemented as standalone websites. 

Details of the baseline system's user interface can be found in Appendix~\ref{sec:baseline}. To ensure that participants fully understand how to use the system, we spent 20 minutes training participants on using it and asked participants to try out the system with test tasks before the formal tasks. It should be noted that while \sys{} offers line correctness, which the baseline system lacks, this is a unique component of our paper. Apart from this, both systems provide similar information. The primary distinction lies in the methods of information presentation. Consequently, we consider our study design to be a fair basis for comparison.

\subsubsection{Programming Problems and Students' Solutions}
To ensure the authenticity of the data used in the study, we collected data from a large introductory programming course at \textit{[redacted for anonymity]}. This data consisted of students' solutions to two distinct programming problems assigned in the course, completed on their own time. The data were collected from an interactive Python textbook used by the course. The data contain genuine examples of mistakes and common patterns students when approaching the problems. To maintain comparability across the systems, we selected one programming exercise from the dataset for each system that had a comparable level of complexity. \add{Specifically, to complete these two exercises, students need to iterate through the provided list, perform string comparison, and then modify the value in other list or dictionary variables.}

Exercise 1 (E1): For each word in \texttt{words}, add `d' to the end of the word if it ends in `e' to make it past tense. Otherwise, add `ed' to make it past tense. Save these past tense words to a list \texttt{past\_tense}.

Exercise 2 (E2): Given a string, return a variable \texttt{counts}, where the keys are letters in the string, the values are how many times each letter appears in the string.

E1 and E2 each had 3496 and 3249 Python code examples. The solutions varied from 3 lines to 15 lines of code. We checked each submission to ensure that it did not contain any identifying information or present any privacy concerns and anonymized appropriately. 

\subsubsection{Study Setup}
The study was conducted remotely using Zoom. Participants joined two sessions, one that used the baseline system and one that used \sys{}. In the first session, participants all worked on quiz questions about E1 and in the second session, participants all worked on quiz questions about E2. 
System order was counterbalanced and we provided 20 minutes of training on how to use the system in each session. 
After training, participants had 20 minutes to answer quiz questions about students' mistakes and approaches using the assigned system. After finishing the quiz questions, participants were asked to complete a survey about their experience using the system. After the second session (with the same procedure but using the other system), we conducted a reflective interview to compare the two systems. Each participant was compensated with a \$25 USD Amazon Gift Card for the completion of each session. 

\subsubsection{Question Design}
To ensure a fair comparison between \sys{} and the baseline system, we carefully designed our quiz questions to include both multiple-choice and open-ended types.
For the multiple-choice questions, we framed them in ways like "Find how many students made this mistake or pattern in their code. Select the closest number from the options below." This approach helped maintain fairness for both systems.
On the other hand, the open-ended questions asked participants to identify as many common mistakes as they could find in students' code using each system.
To establish correct answers for the quiz questions, one of our team members compiled a list of accurate responses. Importantly, we didn't rely directly on the numbers generated by LLMs when creating the answer options. To ensure objectivity, a team member used both \sys{} and the baseline system to perform the tasks and designed the correct options to closely match both conditions. This approach was crucial for maintaining impartiality in our evaluation.

\subsubsection{Data Collection and Metrics}\label{sec:metrics}
During the study, we recorded participants' screens as they performed the tasks, as well as their responses to the quiz questions, their audio think-aloud processes, and their answers to the post-study survey and follow-up interview. For each session, one member of the research team was present.

We used two metrics to evaluate participants' answers to the quiz questions. For the multiple-choice questions, we calculated 
\[
\frac{\text{the number of matched answers}}{max\text{(total number of correct answers, total number of selected answers)}}
\]
to work out their accuracy for the questions.
\add{For a quiz questions that have four options A, B, C, and D, where A and B are correct, if a participant selected A, C, and D, the accuracy is 1 / max(2, 3), which is 0.33. }
For the open-ended questions, we coded the valid mistakes participants found during the study based on a list of existing mistakes generated by the researcher. 

We created a list of code scheme of behaviors observed within the screen recordings. We also coded the screen recordings to analyze the time spent on the multiple-choice questions in the quiz. For the open-ended questions in the quiz, we analyzed the screen recordings to understand how participants interacted with the tool to perform the tasks. For the post-study survey and the follow-up interview data, one member of the research team used a thematic analysis to identify recurring themes and insights in the data. We used a paired t-test for the statistical analysis.

\subsection{Results}\label{sec:results}
We present results of the user study below. We mainly focused on participants' performance in the quiz questions. 

\subsubsection{Accuracy of Multiple-Choice Questions in the Quiz}

As mentioned in Section~\ref{sec:metrics}, we calculated 
\[
\frac{\text{the number of matched answers}}{max\text{(total number of correct answers, total number of selected answers)}}
\]
to work out their accuracy for the multiple-choice questions in the quiz. 
We found that participants using \sys{} ($\mu$ = 93.02, $\sigma$ = 0.06) had accuracy significantly higher than using the baseline system ($\mu$ = 52.40, $\sigma$ = 0.18, $p$ < 0.0001).

\subsubsection{Time to Identify Patterns and Mistakes in Code} 

To understand the influence of the systems on the duration of time participants spent to answer the multiple choice questions, we computed how long it took to answer those questions that related to student mistakes and patterns.
Participants using \sys{} ($\mu$ = 499.06 \add{seconds}, $\sigma$ = 201.47) completed the questions significantly faster then using the baseline system ($\mu$ = 817.50 \add{seconds}, $\sigma$ = 264.85, $p$ < 0.001). One participant mentioned that code clustering at line level visualized the errors clearly in \sys{} (P15).

\subsubsection{Number of Valid Mistakes} 
As mentioned in Section~\ref{sec:metrics}, we coded the valid mistakes participants found in the open-ended questions in the quiz.
The results showed that participants found significantly more valid mistakes using \sys{} ($\mu$ = 4.81, $\sigma$ = 1.91 than the baseline system ($\mu$ = 2.375, $\sigma$ = 1.58, $p$ < 0.001). 
We listed the mistakes participants identified in Appendix (Table~\ref{tab:mistakes-detail-1}).
When using the baseline system, 5 participants only identified syntax errors and described them in a general way, such as ``Type Error'' and ``Name Error'', and 3 participants did not list any mistakes due to time constraints that prevented them from thoroughly reviewing the majority of the solutions. Conversely, when using \sys{}, all 16 participants were able to comprehensively explore the entire solution set and provide detailed descriptions of the identified mistakes by pointing out the areas that made the solution incorrect. 

\subsection{Findings}

% \todo{merge qual and behavior to "findings"}
We analyzed participants' answers to the post-study survey (as shown in Table~\ref{tab:survey_result}) and their interview responses. To understand \sys{}'s usability benefits or issues, we also analyzed participants' interaction and behavior patterns from the video.

\subsubsection{Participants found \sys{} helpful in understanding students' code}

On a scale of 7 where 1 is completely disagree and 7 is completely agree, participants disagreed that \sys{} is less helpful in understanding students' approaches than the baseline system ($\mu$ = 3.38, $\sigma$ = 1.80). 
9 participants (P1-4, P7-10, P13) expressed that \sys{} was more helpful in understanding students' approaches than the baseline system because of the color-coded histogram (P13) and the overall semantic flow (P8, P13). \sys{} highlighted the majority of students and allowed users to layer the filters on code submissions (P13). 1 participant thought \sys{} and the baseline system were comparable in understanding approaches (P14). 
% \participantquote{I think the bars showed like where the majority of students were going and being able to layer the filters and click on the histogram part was really helpful [...]}{P13}

Participants also find it \sys{} more helpful in understanding the code structure than the baseline system. On a scale of 7 where 1 is completely disagree and 7 is completely agree, participants disagreed that \sys{} is less helpful in understanding the code structure information than the baseline system ($\mu$ = 2.94, $\sigma$ = 1.48). 
In the interview responses, 11 participants expressed that they preferred \sys{} in understanding the code structure (P1-5, P7-10, P13, P16). \sys{} presented a concise view that elucidates the primary steps students take in their code, thereby showing how most students structure their code to solve a programming exercise (P1-2, P5, P7-8). In contrast, the baseline system required users to read every individual code solution and search for a specific pattern of a few lines of code, taking more effort (P2-3, P7, P10-11).

However, some participants also mentioned that the grid view in the baseline system is simple and straightforward, and had lower learning curve for them (P5-6, P11-12, P15-16)

% reason

\subsubsection{\sys{} fostered an exploratory and brainstorming approach to understand students' code}
To better understand how participants use both systems to comprehend students' code, we analyzed their interaction with the systems. We found that participants had very different strategies in answering questions across the two system. 

% what are the different strategies?
% \add{add more details to support the claim?}
In the baseline system, participants had two strategies, 1) random browsing and 2) active searching without guidance. Since the baseline system directly displayed clusters derived from \overcode{}, each incorrect solution was categorized into its own cluster. Without categorized information of mistakes, some participants tried to look through all the incorrect solutions and reported mistakes they encountered during the process (P1-4, P7, P9, P15). 
Other participants would first come up with some coding patterns and search for them, and then look for other unseen patterns by filtering out what has been seen (P10, P13). Usually what they come up with is a correct pattern instead of incorrect patterns. 
Due to the lack of guidance, browsing mistakes using the baseline system largely depended on what users had in mind and required users to have an expectation about potential mistakes students would have.

In contrast, when using \sys{}, all participants first looked at the semantic-label abstractions and the color-coded histogram to locate where most mistakes were and what the common mistakes were. With the guidance provided in \sys{}, they then selectively dug deeper into the details to reason about students' misconceptions. 

The difference indicated that \sys{} and the baseline system each better fits different settings. participants expressed a preference for using \sys{} when identifying common mistakes and exploring students' code on a larger scale, particularly in extensive programming classes with hundreds or thousands of students (P2, P7-8, P10-11, P13). 
The efficient navigation capabilities in \sys{} enables swift exploration of students' code (P2, P5-7), while the baseline system required participants to read each code solution individually (P1, P3, P5, P7, P9-10, P13, P15).
Some participants expressed a preference for the baseline system during smaller sessions when seeking specific patterns (P1, P8, P10, P13).

% We also found that whether being able to fully explore and brainstorm could affect how participants comprehend students' code. 
% For example, when using the baseline system to answer quiz questions, some participants mentioned they could not get an accurate number from the system, and they had to select an answer based on their previous experience and impression of the programming exercise (P1-2, P5, P9, P11).

Furthermore, we found that the ability to explore and brainstorm students' code might be beneficial to people with little teaching experience.
We analyzed the mistakes listed by two participants who did not have teaching experience. P10 did not have any time to list common mistakes when using the baseline system and listed 4 mistakes with details when using \sys{}. P6 listed only 4 general types of syntax errors when using the baseline system, while in \sys{} listed 7 mistakes with details including semantic errors. This being said, with \sys{}'s guidance in browsing mistakes, even participants with limited teaching experience and few expectations of students' code could effectively identify common mistakes.

\subsubsection{\sys{} help participants navigate students' solutions with less context switching}

Based on our observation, we found that \sys{} takes participants less context switch to understand mistakes within thousands of code solutions.
In \sys{}, participants can quickly understand errors of the whole classroom by looking at the dominant patterns in SAV and SHV, and inspecting on individual solutions by clicking to view CDV. 
Participants noted that they could effortlessly locate the information they desired and seamlessly switch between abstraction and finer details (P1-3, P5, P9-10, P12-13, P15).

Participants were asked to identify common implementations used by students. When using the baseline system, 4 out of 8 participants generated search queries, filtered out relevant portions of the dataset, and repeated this process until they no longer observed common implementations. On the other hand, 4 participants listed a single implementation and moved on to other questions. However, when utilizing \sys{}, all participants initially located a specific step within the semantic flow, expanded the step to view all implementations, and identified common implementations by referring to the numerical counts alongside them in the histogram.

When locating the steps that have the most mistakes in \sys{}, all participants directly used the color-code histogram. They looked at both the dominant color of the histogram and the point where it transitioned from green to red. They then inferred that the step where the majority of it was light green should have had more mistakes than those where the majority of it was dark green.

% \subsubsection{Visualizing students' solutions at scale still can be overwhelming in both \sys{} and the baseline system}

% Despite the benefits mentioned above, 4 participants found the code semantic view in \sys{} to be overwhelming when they expand it to show all the implementations. While \sys{} concisely visualized a large set of solutions that looked like a simple program, they mentioned that they needed to explore and click to check code details, during which they lost track of their clicking history and where they were at (P5, P10-11, P16).

% While most participants found the baseline system's grid view to be overwhelming as well, and not as easy to navigate as \sys{} (P1-3, P8, P10, P15-16), some participants stated that the grid view saved them from clicking and did not require them to switch between different levels of representation and keep track of where they were (P5, P11). 

\section{discussion}

% This paper introduced the \sys{} system and demonstrated how its semantic flow view design  facilitated the visualization and comprehension of student code on a large scale. 
% We explored a novel design  by implementing the following key features:
% \begin{itemize}
%     \item \textbf{Concise Presentation}: It condensed thousands of students' solutions into a streamlined view that adhered to a common semantic flow.
%     \item \textbf{Line-Level Nuance}: It visualized subtle differences at the code line level by associating code lines with steps in the semantic flow.
%     \item \textbf{Semantic Enrichment}: It enriched a visualization's semantic meaning by incorporating representative code lines from various implementations into the semantic flow view.
%     \item \textbf{Distribution Insights}: It offered a preview of the distribution of various implementations and their correctness through the use of color-coded histograms.
% \end{itemize}

In this section, we (1) discuss how \sys{} contributes to the body of work in visualizing students' code at scale, (2) discuss the role of LLM in \sys{}, and (3) discuss \sys{}'s limitation.

\subsection{Connecting Individual Code Solutions with the Broad Context of the Whole Dataset}

Our findings highlight the transformative impact of \sys{}. 
By compacting thousands of student solutions into a singular, coherent view that traces a shared semantic trajectory, \sys{} unlocked how participants interpreted and interacted with the data from two perspectives: they could immerse themselves in the specifics of individual solutions, yet never lose sight of the grand tapestry of submissions. The color-coded error information on \sys{}'s semantic code view enriched this experience, illuminating subtle line-level differences and fostering a more nuanced understanding.

In contrast, the baseline system required participants to inspect each incorrect solution individually to identify errors and search for specific patterns. This approach was less integrated than the aggregated view offered by \sys{}, which naturally linked individual solutions to the broader dataset (as noted by P1-4, P7, P9, P15). While tools like \overcode{} reduced the number of solutions instructors had to review, their grid layouts often isolated solutions. In comparison, \sys{} effectively demonstrated the connections between individual solutions.

\subsection{Bridging Abstraction and Code Details}

When viewing code solutions at scale, users often toggle between two distinct levels of abstraction: the high-level overview of the entire solution set and the low-level details of individual solutions. Our user study revealed that \sys{} introduces an additional intermediate level of information, effectively bridging the gap between the high-level abstractions and the details of code. 
For instance, by looking at the initial visualization without interaction, one participant noticed a reduction in the histogram's size after variable initialization and for loops. This indicated that students might encounter more challenges when writing conditional statements, as many submissions ended at the content inside the for loops. Furthermore, we found that participants could identify common approaches through the preview of the code lines associated with the visualization, without delving into the exact code content.

\sys{} offers a novel approach for effectively viewing extensive code data. While many existing tools have introduced innovative presentations, such as \vizprog{}'s dynamic dots on a 2D map~\cite{zhang2023vizprog}, they often fall short in conveying semantic meaning through their visual positioning. This absence forces instructors to manually select specific areas and delve into the raw code to understand the content. \sys{} pioneers a design that seamlessly integrates code lines into a semantic view, replete with visual cues. This approach underscores the advantages of presenting multiple levels of code representation, effectively bridging the gap seen in previous tools.

\begin{figure}
    \centering
    \includegraphics[width=0.5\textwidth]{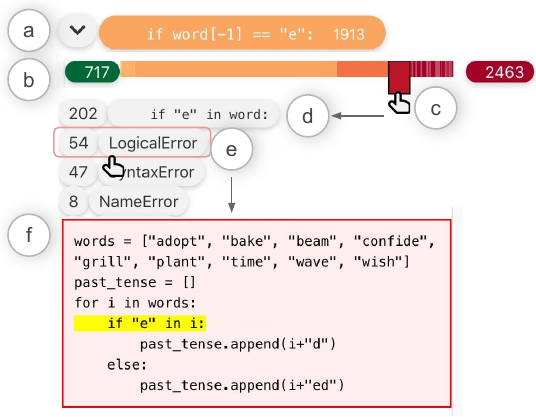}
    \vspace{-0.2in}
    \caption{An example of how \sys{} looks like without LLM determining line correctness. (a) is a collection of code lines that check the end of a word, and (b) is the correctness histogram. Upon selecting the prominent red block (c), users can view an example that incorrectly check the end of a word (d). By clicking on ``LogicalError'' (e), users are then able to explore detailed solutions (f).}
    \label{fig:no-llm}
    \vspace{-0.2in}
\end{figure}

% \vspace{-0.2in}

\add{\subsection{LLM's Role in \sys{}}\label{sec:llm-discussion}
\sys{} uses LLM to generate line-level error information for student code. We discuss LLM's role in \sys{} from two angles: 1) the accuracy of LLM, and 2) \sys{}'s dependency on LLM.}

\sys{} is primarily designed to provide instructors with an overarching view of the entire solution set to facilitate exploration and analysis. Therefore, the LLM does not need to produce perfectly accurate error information for each code line. Our user study revealed that participants primarily use the color distribution in the histogram view to pinpoint steps where most students have mistakes, and then examine the specific code lines for a deeper understanding of these errors. 
Even if disregarding thel ine correctness information from LLM and labeling all lines from an incorrect solution as erroneous, \sys{}'s visual design could provide valuable insights into students' mistakes. For instance, Figure~\ref{fig:no-llm} is an example without LLM's outputs, where instructors can still observe code lines grouped by semantic meanings (Figure~\ref{fig:no-llm}a) and color-coded for correctness (Figure~\ref{fig:no-llm}b). After clicking a prominent red block (Figure~\ref{fig:no-llm}c), instructors can see that the cluster represents the code line ``\texttt{if "e" in word:}'' (Figure~\ref{fig:no-llm}d), which incorrectly check the end of a word. Upon clicking ``LogicalError'' (Figure~\ref{fig:no-llm}e), instructors can then explore detailed solutions (Figure~\ref{fig:no-llm}f). 
Despite the less pronounced color difference compared to the original design, instructors are still able to get useful information and navigate students' solutions.

While \sys{} relies on LLM for locating semantic errors, we argue that LLM can be replaced in \sys{}. Any tool capable of pinpointing semantic errors could potentially replace LLM in \sys{}. Alternatives might include a smaller language model or an earlier version of ChatGPT. Moreover, implementing more targeted test cases could help in detecting the precise lines with errors. We chose ChatGPT-3.5 for its widespread accessibility and ease of use. For real classroom deployment, instructors have the flexibility to substitute LLM with any other tool that effectively locates semantic errors.

\subsection{Limitations and Future Work}\label{sec:limitation-llm-reliability}
One limitation is the costs associated with LLM usage and its generalizability to real-time settings. \sys{} integrated an LLM as a post-hoc code analysis tool, while in real-time usage, costs associated with doing so can accumulate significantly, thus impacting both the financial and time resources required to maintain the system. Therefore, future work should explore alternatives that balance computational demands with real-time usage needs.

% \gez{removed "Feature Refinement"}

\section{conclusion}

% In this work, we explored a design that allows instructors to visualize and understand the semantic flow in student's code at scale for programming exercises. 
% We introduce \sys{}, a tool that allow instructors to explore large numbers of students' coding submissions with 

In this research, we addressed the challenges faced by instructors when analyzing large numbers of varied code submissions from students.
We introduced \sys{}, a novel system that uses semantic labeling and code structure to visualize the semantic flow of students' submissions. 
By abstracting code statements based on their meaning while maintaining structure, \sys{} offers a comprehensive view that simplifies navigation and comparison of code flows. 
Our evaluation showed that \sys{} allows educators to analyze these submissions more effectively than traditional methods. Participants found it easier to explore patterns and understand code structure with \sys{}. This work highlights the potential for improving teaching methods by better understanding student submissions and delivering targeted feedback at scale.

\begin{acks}
To Robert, for the bagels and explaining CMYK and color spaces.
\end{acks}

\bibliographystyle{ACM-Reference-Format}
\bibliography{reference}

% \appendix

\newpage
\appendix
\section{Mistakes participants found in open-ended questions}

\begin{table*}[]
    \centering
    \small
    \resizebox{\textwidth}{!}{
\begin{tabular} [c]{c p{1.5cm} l l }
    \hline
        \textbf{PID} & \textbf{Condition} & \textbf{Mistakes} & \textbf{Count} \\
    \hline
        P1 & \baseline{} & None & 0 \\ 
        & \sys{} & mix up variable names, looping over empty dictionary instead of the word, use variables that are not defined, initialize as 0 for the first time encountered ant not adding 1 , looping over the keywords of an empty dictionary" & 5 \\ \hline
        
        P2 & \baseline{} & concatenate an append statement to a str variable, missing quotation marks around ``d'' and ``ed'', use wrong conditional statement for modifying the word" & 3 \\
        & \sys{} & mix up variable name, syntax error in writing conditions, not referring to the ``counts'' variable correctly, incorrect use of the ``str.split() method'' & 4 \\ \hline
        
        P3 & \baseline{} & indentation error & 1 \\ 
        & \sys{} & syntax error in writing condition, use variables that are not defined, using ``is'' instead of ``=='' when checking str values, use ``()'' to index dictionary instead of ``[]'' & 4 \\ \hline

        P4 & \baseline{} & not define past\_tense, syntax error on adding ``d'' or ``ed'' & 2 \\ 
        & \sys{} & looping over empty dictionary instead of the word, syntax error in writing conditions, use variables that are not defined, Key Error, indentation error, type error, attribute error, using str.split() when it is not necessary, mix up variable names, missing colon in for loop" & 10 \\ \hline

        P5 & \baseline{} & Name Error, Key Error, Logical Error, type error, syntax error in wriitng conditions, loop over wrong data type & 6 \\ 
        & \sys{} & use append method on a str variable, incorrectly format append, modify the word but not assign it to any variable, concatenate an append statement to a str variable & 4 \\ \hline

        P6 & \baseline{} & Name Error, Key Error, Logical Error, Syntax Error & 4 \\ & \sys{} & modify the word but not assign it to any variable, add quotation marks around variable names, append ``d'' to the list instead of the modified word, add ``d'' to word[-1] instead of word, use ``=='' instead ``='' to assign values, trying to reassign the append statement to the list, use wrong conditional statement for modifying the word & 7 \\ 
        \hline

        P7  & \baseline{} & mix up variable names, missing loops, use variables that are not defined & 3 \\
        & \sys{} & "append ``d'' to the list instead of the modified word, attribute error, syntax error, modify the word but not assign it to any variable, modify the word but not append it to the list, use wrong index when checking the last letter of word"  & 6 \\  
        \hline

        P8 & \baseline{} & use append method on a str variable, modify the word but not assign it to any variable, use wrong index when checking the last letter of word" & 3 \\ 
        & \sys{} & syntax error in writing conditions, do not understand how loop over str works & 2 \\ \hline

        P9 & \baseline{} & None & 0 \\ 
        & \sys{} & indentation error, missing content after for loop & 2 \\ \hline

        P10 & \baseline{} & None & 0 \\ 
        & \sys{} & modify dictionary values without initializing it, mix up variable names, looping over empty dictionary instead of the word, use variables that are not defined & 4 \\ \hline

        P11 & \baseline{} & Name Error, Key Error, incorrectly index letters in str variable & 3 \\
        & \sys{} & incorrectly format append, compare a str variable with a list using ``<'', overwriting the list with a word instead of updating the variable, updating the index instead of the word, initialize list with a str variable & 5 \\ 
         \hline

        P12 & \baseline{} & check the whole word instead of the letter in conditional statements, use variables that are not defined, mix up variable names & 3 \\
        & \sys{} &  append ``d'' to the list instead of the modified word, use append method on a str variable, use variable undefined, use wrong conditional statement for modifying the word, use ``='' instead of ``=='' when checking the end of word, indentation error & 6 \\ 
         \hline

        P13 & \baseline{} & modify the word but not append it to the list & 1 \\ 
        & \sys{} & mix up variable names, indentation error, modify dictionary values without initializing it, check if letter is in the original str variable instead of the new dictionary & 4 \\ \hline

        P14 & \baseline{} & missing conditional statement, Logical Error, wrong conditional statement for initializing dictionary & 3 \\
        & \sys{} & append ``d'' to the list instead of the modified word, modify the word but not assign it to any variable, modify the original list ``word'' but not creating a new list ``past\_tense'' & 3 \\
         \hline

        P15 & \baseline{} & mix up variable names, check if letter is in the original str variable instead of the new dictionary, indentation error" & 3 \\
        & \sys{} & use append method on a str variable, use the ``join'' method instead of ``append'', incorrectly format append, trying to reassign the append statement to the list, concatenate an append statement to a str variable, missing content inside the loop & 6 \\ 
         \hline

        P16 & \baseline{} & use variables that are not defined, wrong conditional statement for initializing dictionary, loop over wrong data type & 3 \\
        & \sys{} & not change the word before append it to list, use ````e'' in word'' to check the end of a word, indentation error, directly modify word without conditional statements, syntax error on adding ``d'' or ``ed'' & 5 \\ 
         \hline

        \hline
    % \caption{Valid mistakes participants found in open-ended quiz quesitons.}\label{tab:mistakes-detail-1}\\

\end{tabular}
}
    \caption{Mistakes participants found in open-ended questions}
    \label{tab:mistakes-detail-1}
\end{table*}

\section{Post-study survey questions}

\begin{table*}[]
    \centering
    \footnotesize
    \begin{tabular}{p{5cm}cccc}
        \toprule
         \textbf{Statement} & \textbf{Mean} & \textbf{SD} & \textbf{Median} & \textbf{Iqr} \\
         \\
         \midrule
        \multirow{2}{=}{S1: Compared to \runex{}, \sys{} takes less time to identify students' misconception. } & 6.06 & 1.48 & 7.00 & 1.00\\
        & \\
        \midrule
        \multirow{2}{=}{S2: Compared to \runex{}, \sys{} is less helpful in understanding students' approaches. } & 3.38 & 1.80 & 2.50 & 3.00 \\
        & \\
        \midrule
        \multirow{2}{=}{S3: Compared to \runex{}, \sys{} better supported me to explore students' solutions. } & 4.63 & 1.62 & 5.00 & 3.00\\
        & \\
        \midrule
        \multirow{2}{=}{S4: Compared to \runex{}, \sys{} is less helpful in understanding the code structure information.} & 2.94 & 1.48 & 3.00 & 2.00\\
        & \\
        \midrule
        \multirow{2}{=}{S5: Compared to \runex{}, I would prefer to use \sys{} for similar tasks in the future. } & 5.63 & 1.73 & 6.00 & 2.00 \\
        & \\
        \bottomrule
    \end{tabular}
    \caption{Post-study survey results of comparing \sys{} and the baseline system. Participants were asked to rate their agreement with various statements on a scale from 1 to 7, where 1 indicated complete disagreement. To ensure impartiality, we replaced the term \baseline{} with \runex{} to prevent participants from discerning which system was the baseline.}
    \label{tab:survey_result}
\end{table*}

\section{Baseline System}\label{sec:baseline}
\begin{figure*}
    \centering
    \includegraphics[width=\textwidth]{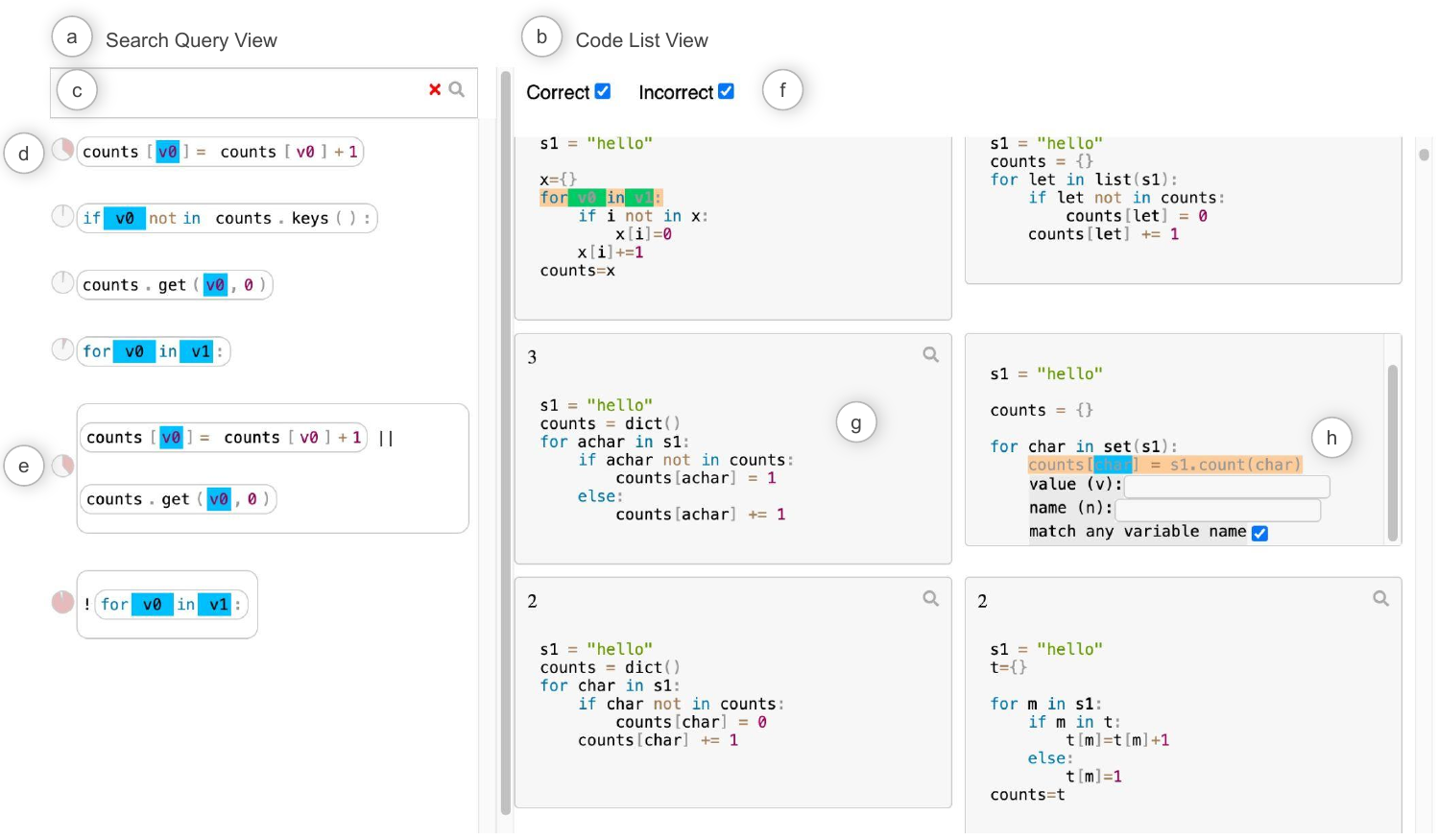}
    \caption{\add{The baseline system's user interface, derived from \runex{} and incorporating \overcode{}'s clustering results, features two main views: the Search Query View (a) and the Code List View (b). In the Code List View, each code block (g) represents a cluster of solutions with identical computation results, with a number in the upper left corner indicating the cluster size. Users can search for specific code patterns using runtime values and text matching (h), with each query displayed in the Search Query View alongside descriptive statistics (d). Queries can be entered directly into the search bar (c). Additionally, the system allows for set operations on these queries (e). Users can filter the code blocks in the Code List View by clicking on a query or using the checkbox (f). } }
    \label{fig:baseline}
\end{figure*}

Figure \ref{fig:baseline} shows the baseline system's user interface, which is derived from \runex{} and incorporating \overcode{}’s clustering results. Each code block (Figure~\ref{fig:baseline}g) represents either a cluster of correct solutions, or a single incorrect solutions, as computed by \overcode{}. To maintain a fair comparison, we take the exact same prototype from \runex{}~\cite{zhang2023runex}, except for that here each code block represents a cluster of solutions, while in the original prototype each code block represents a single solution. In the baseline system's user interface, code blocks are displayed in a grid view. In addition, users can easily create search queries in the baseline system by simply selecting code lines and typing in conditions to search through the entire solution set (Figure~\ref{fig:baseline}h). Each query is displayed alongside with descriptive statistics (Figure~\ref{fig:baseline}d). The systems allows for set operations on the queries (Figure~\ref{fig:baseline}e). The baseline system supports participants both view students' solutions in clusters and search through the whole solution set with text matching and runtime values. Users can easily filter the dataset through clicking the queries and the check box (Figure~\ref{fig:baseline}f). A thorough evaluation on the usability and effectiveness of the baseline system can be found in Zhang et al.'s paper~\cite{zhang2023runex}. 

\section{Prompt}\label{sec:prompt}
We used the following prompt for the determination:
\begin{quote}
    Prompt: \\
    \textquotesingle\textquotesingle\textquotesingle\\
     Imagine you are a Python tutor, and there are some code samples for you to review. For each code sample below, there are some Attribute/Syntax/Value/Name/Type/Logical Errors. Read through it line by line and identify Errors based on the original Compile Error and the problem description given below. If you find out an Error, please do double check by reading codes around it. Be really careful while dealing with variables' names.\\
Output all errors and keep the error description short.\\
The output must strictly follow the Output Format below.\\
Problem Description:\\
\{Problem Description\}\\

Output Format:\\
\textasciigrave\textasciigrave\textasciigrave\\
ERROR \#: ERROR TYPE: ERROR Description  |  Line Number \#: "Code"\\

ERROR \#: ERROR TYPE: ERROR Description  |  Line Number \#: "Code"\\
\textasciigrave\textasciigrave\textasciigrave\\
Code Sample Example:\\
\textasciigrave\textasciigrave\textasciigrave\\
\{Input Example\}\\
\textasciigrave\textasciigrave\textasciigrave\\
Output Example:\\
\textasciigrave\textasciigrave\textasciigrave\\
\{Output Example\}\\
\textasciigrave\textasciigrave\textasciigrave\\
(Line Number starts from 0. Stop when there is no error left. No more than 7 ERRORS!! Make sure that there is no repeated ERROR!!)\\

Think step by step, and do self-reflection and self-correction before output your answer! Do double check across the code lines if you think you found a variable that is not defined!!!\\

Original Compile Error : \{OEmessage\}\\
\textquotesingle\textquotesingle\textquotesingle
% \yc{fill in the details, @Ashley}
\end{quote}

\end{document}